\documentclass[a4paper,oneside,12pt,draftcls,journal,onecolumn]{IEEEtran}%
\usepackage{amsfonts}
\usepackage{amsmath}
\usepackage{amssymb}
\usepackage{graphicx}%
\providecommand{\U}[1]{\protect\rule{.1in}{.1in}}

\begin{document}

\title{Additive-State-Decomposition-Based Tracking Control for TORA Benchmark}
\author{Quan Quan and Kai-Yuan Cai \thanks{The authors are with Department of
Automatic Control, Beijing University of Aeronautics and Astronautics, Beijing
100191, China. Q. Quan, Assistant Professor (Lecturer), qq\_buaa@buaa.edu.cn,
http://quanquan.buaa.edu.cn. K.-Y. Cai, Professor, kycai@buaa.edu.cn.}}
\maketitle

\begin{abstract}
In this paper, a new control scheme, called
\emph{additive-state-decomposition-based tracking control}, is proposed to
solve the tracking (rejection) problem for rotational position of the TORA (a
nonlinear nonminimum phase system). By the additive state decomposition, the
tracking (rejection) task for the considered nonlinear system is decomposed
into two independent subtasks: a tracking (rejection) subtask for a linear
time invariant (LTI) system, leaving a stabilization subtask for a derived
nonlinear system. By the decomposition, the proposed tracking control scheme
avoids solving regulation equations and can tackle the tracking (rejection)
problem in the presence of any external signal (except for the frequencies at
$\pm1$) generated by a marginally stable autonomous LTI system. To demonstrate
the effectiveness, numerical simulation is given.

\end{abstract}


\section{Introduction}

The tracking (rejection) problem for a nonlinear benchmark system called
translational oscillator with a rotational actuator (TORA) and also known as
rotational-translational actuator (RTAC) has received a considerable amount of
attention these years \cite{Wan (1996)}-\cite{Jiang(2009)}. Some results were
presented concerning the tracking (rejection) problem for general external
signals \cite{Zhao(1998)},\cite{Jiang(2000)}. However, the proposed control
methods cannot achieve asymptotic disturbance rejection. Taking this into
account, the nonlinear output regulation theory was applied to track (reject)
external signals generated by an autonomous system. In this case, asymptotic
disturbance rejection can be achieved. By using different measurement, the
tracking (rejection) problem for translational displacement of the TORA were
investigated \cite{Huang(2004)}-\cite{Fabio(2011)}. Readers can refer to
\cite{Fabio(2011)} for details. Based on the same benchmark system, some other
work was also presented concerning the tracking (rejection) problem for
rotational position by nonlinear output regulation theory \cite{Lan(2006)}%
,\cite{Jiang(2009)}. For the two types of tracking (rejection) problems,
regulator equations have to be solved and then the resulting solutions will be
further used in the controller design. However, the difficulty of constructing
and solving regulator equations will increase as the complexity of external
signals increases. Moreover, it may fail to design a controller if regulator
equations have no solutions. These are our major motivation.

In this paper, the tracking (rejection) problem for rotational position of the
TORA as \cite{Lan(2006)},\cite{Jiang(2009)} is revisited by a new control
scheme called \emph{additive-state-decomposition-based tracking control},
which is based on the \emph{additive state decomposition}\footnote{In this
paper we have replaced the term \textquotedblleft additive
decomposition\textquotedblright\ in \cite{Quan(2009)} with the more
descriptive term \textquotedblleft additive state
decomposition\textquotedblright.}. The proposed additive state decomposition
is a new decomposition manner different from the lower-order subsystem
decomposition methods. Concretely, taking the system $\dot{x}\left(  t\right)
=f\left(  t,x\right)  ,x\in%
\mathbb{R}
^{n}$ for example, it is decomposed into two subsystems: $\dot{x}_{1}\left(
t\right)  =f_{1}\left(  t,x_{1},x_{2}\right)  $ and $\dot{x}_{2}\left(
t\right)  =f_{2}\left(  t,x_{1},x_{2}\right)  $, where $x_{1}\in%
\mathbb{R}
^{n_{1}}\ $and$\ x_{2}\in%
\mathbb{R}
^{n_{2}},$ respectively. The lower-order subsystem decomposition satisfies%
\[
n=n_{1}+n_{2}\text{ and }x=x_{1}\oplus x_{2}.
\]
By contrast, the proposed additive state decomposition satisfies%
\[
n=n_{1}=n_{2}\text{ and}\ x=x_{1}+x_{2}.
\]
In our opinion, lower-order subsystem decomposition aims to reduce the
complexity of the system itself, while the additive state decomposition
emphasizes the reduction of the complexity of tasks for the system.

By following the philosophy above, the original tracking (rejection) task is
`additively' decomposed into two independent subtasks, namely the tracking
(rejection) subtask for a linear time invariant (LTI) system and the
stabilization subtask for a derived nonlinear system. Since tracking
(rejection) subtask only needs to be achieved on an LTI system, the complexity
of external signals can be handled easier by the transfer function method. It
is proved that the designed controller can tackle the tracking (rejection)
problem for rotational position of the TORA in the presence of any external
signal (except for the frequency at $\pm1$) generated by a marginally stable
autonomous LTI system.

This paper is organized as follows. In Section 2, the problem is formulated
and the additive state decomposition is recalled briefly first. In Section 3,
an observer is proposed to compensate for nonlinearity; then the resulting
system is `additively' decomposed into two subsystems; sequently, controllers
are designed for them. In Section 4, numerical simulation is given. Section 5
concludes this paper.

\section{Nonlinear Benchmark Problem and Additive State Decomposition}

\subsection{Nonlinear Benchmark Problem}

As shown in Fig.1, the TORA system consists of a cart attached to a wall with
a spring. The cart is affected by a disturbance force $F$. An unbalanced point
mass rotates around the axis in the center of the cart, which is actuated by a
control torque $N.$ The translational displacement of the cart is denoted by
$x_{c}$ and the rotational position of the unbalanced point mass is denoted by
$\theta.$

\begin{figure}[h]
\begin{center}
\includegraphics[
scale=1.3]{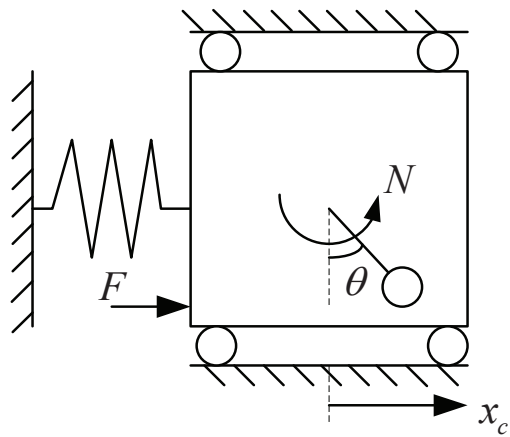}
\end{center}
\caption{TORA system configuration \cite{Wan (1996)}}%
\end{figure}

For simplicity, after normalization and transformation, the TORA system is
described by the following state-space representation \cite{Wan (1996)}:
\begin{subequations}
\label{RTACmodel0}%
\begin{align}
\dot{x}_{1}  &  =x_{2}\label{RTACmodel1}\\
\dot{x}_{2}  &  =-x_{1}+\varepsilon\sin x_{3}+F_{d}\label{RTACmodel2}\\
\dot{x}_{3}  &  =x_{4}\label{RTACmodel3}\\
\dot{x}_{4}  &  =u-\frac{\varepsilon\cos x_{3}}{1-\varepsilon^{2}\cos^{2}%
x_{3}}F_{d},x\left(  0\right)  =x_{0} \label{RTACmodel4}%
\end{align}
where $0<\varepsilon<1,$ $x=[%
\begin{array}
[c]{cccc}%
x_{1} & x_{2} & x_{3} & x_{4}%
\end{array}
]^{T}\in%
\mathbb{R}
^{4}$, $x_{3}=\theta$, $F_{d}\in%
\mathbb{R}
$ is the unknown dimensionless disturbance, $u\in%
\mathbb{R}
$ is the dimensionless control torque. In this paper, the tracking (rejection)
problem for rotational position of the TORA as \cite{Lan(2006)}%
,\cite{Jiang(2009)} is revisited. Concretely, for system (\ref{RTACmodel0}),
it is to design a controller $u$ such that the output $y\left(  t\right)
=x_{3}\left(  t\right)  \rightarrow r$ as $t\rightarrow\infty,$ meanwhile
keeping the other states bounded, where $r\in(-\pi\left/  2\right.
,\pi\left/  2\right.  )$ is a known constant. Obviously, this is a nonlinear
nonminimum phase tracking problem, or say a nonlinear weakly minimum phase
tracking problem. For system (\ref{RTACmodel0}), the following assumptions are imposed.

\textit{Assumption 1}. The state $x$ can be obtained.

\textit{Assumption 2}. The disturbance $F_{d}\in%
\mathbb{R}
$ is generated by an autonomous LTI system%
\end{subequations}
\begin{equation}
\dot{w}=Sw,F_{d}=C_{d}^{T}w \label{RTACdisturbance}%
\end{equation}
where $S=-S^{T}\in%
\mathbb{R}
^{m\times m}$, $C_{d}\in%
\mathbb{R}
^{m}$ are constant matrix, $w\in%
\mathbb{R}
^{m},$ and the pair $\left(  C_{d}^{T},S\right)  $ is observable.

\textit{Remark 1}. If all eigenvalues of $S$ have zero real part, then, in
suitable coordinates, the matrix $S$ can always be written to be a
skew-symmetric matrix. The matrix $S$ in previous literature on the output
regulation problem is often chosen in a simple form $S=\left[
\begin{array}
[c]{cc}%
0 & \omega\\
-\omega & 0
\end{array}
\right]  ,$ where $\omega$ is a positive real \cite{Huang(2004)}%
-\cite{Jiang(2009)}. In such a case, $F_{d}$ is in the form as sin$\left(
\pm\omega t\right)  $ and the solution to the regulator equation is easier to
obtain. However, this is a difficulty when $S$ is complicated.

\subsection{Additive State Decomposition}

In order to make the paper self-contained, the additive state decomposition
\cite{Quan(2009)} is recalled here briefly. Consider the following `original'
system:%
\begin{equation}
f\left(  {t,\dot{x},x}\right)  =0,x\left(  0\right)  =x_{0}
\label{Gen_Orig_Sys}%
\end{equation}
where $x\in%
\mathbb{R}
^{n}$. We first bring in a `primary' system having the same dimension as
(\ref{Gen_Orig_Sys}), according to:%
\begin{equation}
f_{p}\left(  {t,\dot{x}_{p},x_{p}}\right)  =0,x{_{p}}\left(  0\right)
=x_{p,0} \label{Gen_Pri_Sys}%
\end{equation}
where ${x_{p}}\in%
\mathbb{R}
^{n}$. From the original system (\ref{Gen_Orig_Sys}) and the primary system
(\ref{Gen_Pri_Sys}) we derive the following `secondary' system:%
\begin{equation}
f\left(  {t,\dot{x},x}\right)  -f_{p}\left(  {t,\dot{x}_{p},x_{p}}\right)
=0,x\left(  0\right)  =x_{0} \label{Gen_Sec_Sys0}%
\end{equation}
where ${x_{p}}\in%
\mathbb{R}
^{n}$ is given by the primary system (\ref{Gen_Pri_Sys}). Define a new
variable ${x_{s}}\in%
\mathbb{R}
^{n}$ as follows:%
\begin{equation}
{x_{s}\triangleq x-x_{p}}. \label{Gen_RelationPS}%
\end{equation}
Then the secondary system (\ref{Gen_Sec_Sys0}) can be further written as
follows:%
\begin{equation}
f\left(  {t,\dot{x}_{s}+\dot{x}_{p},x_{s}+x_{p}}\right)  -f_{p}\left(
{t,\dot{x}_{p},x_{p}}\right)  =0,x{_{s}}\left(  0\right)  =x_{0}-x_{p,0}.
\label{Gen_Sec_Sys}%
\end{equation}
From the definition (\ref{Gen_RelationPS}), we have%
\begin{equation}
{x}\left(  t\right)  ={x_{p}\left(  t\right)  +x_{s}\left(  t\right)
,t\geq0.} \label{Gen_RelationPS1}%
\end{equation}

\textit{Remark 2.} By the additive state decomposition\textit{, }%
the\textit{\ }system (\ref{Gen_Orig_Sys}) is decomposed into two subsystems
with the same dimension as the original system. In this sense our
decomposition is \textquotedblleft additive\textquotedblright. In addition,
this decomposition is with respect to state. So, we call it \textquotedblleft
additive state decomposition\textquotedblright\emph{.}

As a special case of (\ref{Gen_Orig_Sys}), a class of differential dynamic
systems is considered as follows:%
\begin{align}
\dot{x}  &  =f\left(  {t,x}\right)  ,x\left(  0\right)  =x_{0},\nonumber\\
y  &  =h\left(  {t,x}\right)  \label{Dif_Orig_Sys}%
\end{align}
where ${x}\in%
\mathbb{R}
^{n}$ and $y\in%
\mathbb{R}
^{m}.$ Two systems, denoted by the primary system and (derived) secondary
system respectively, are defined as follows:%
\begin{align}
\dot{x}_{p}  &  =f_{p}\left(  {t,x_{p}}\right)  ,x_{p}\left(  0\right)
=x_{p,0}\nonumber\\
y_{p}  &  =h_{p}\left(  {t,x}_{p}\right)  \label{Dif_Pri_Sys}%
\end{align}
and%
\begin{align}
\dot{x}_{s}  &  =f\left(  {t,x_{p}}+{x_{s}}\right)  -f_{p}\left(  {t,x_{p}%
}\right)  ,x_{s}\left(  0\right)  =x_{0}-x_{p,0},\nonumber\\
y_{s}  &  =h\left(  {t,x_{p}}+{x_{s}}\right)  -h_{p}\left(  {t,x}_{p}\right)
\label{Dif_Sec_Sys}%
\end{align}
where ${x_{s}}\triangleq{x-x_{p}}$ and $y_{s}\triangleq{y-y_{p}}$. The
secondary system (\ref{Dif_Sec_Sys}) is determined by the original system
(\ref{Dif_Orig_Sys}) and the primary system (\ref{Dif_Pri_Sys}). From the
definition, we have%
\begin{equation}
{x}\left(  t\right)  ={x_{p}\left(  t\right)  +x_{s}\left(  t\right)
,y\left(  t\right)  =y_{p}\left(  t\right)  +y_{s}\left(  t\right)  ,t\geq0.}
\label{Gen_RelationDif}%
\end{equation}

\section{Additive-State-Decomposition-Based Tracking Control}

In this section, in order to decrease nonlinearity, an observer is proposed to
compensate for the nonlinear term $\frac{\varepsilon\cos x_{3}}{1-\varepsilon
^{2}\cos^{2}x_{3}}F_{d}.$ After the compensation, the resulting nonlinear
nonminimum phase tracking system\ is decomposed into two systems by the
additive state decomposition: an LTI system including all external signals as
the primary system, leaving the secondary system with a zero equilibrium
point. Therefore, the tracking problem for the original system is
correspondingly decomposed into two subproblems by the additive state
decomposition: a tracking problem for the LTI `primary' system and a
stabilization problem for the secondary system. Obviously, the two subproblems
are easier than the original one. Therefore, the original tracking problem is
simplified. The structure of the closed-loop system is shown in
Fig.2.\begin{figure}[h]
\begin{center}
\includegraphics[
scale=0.8 ]{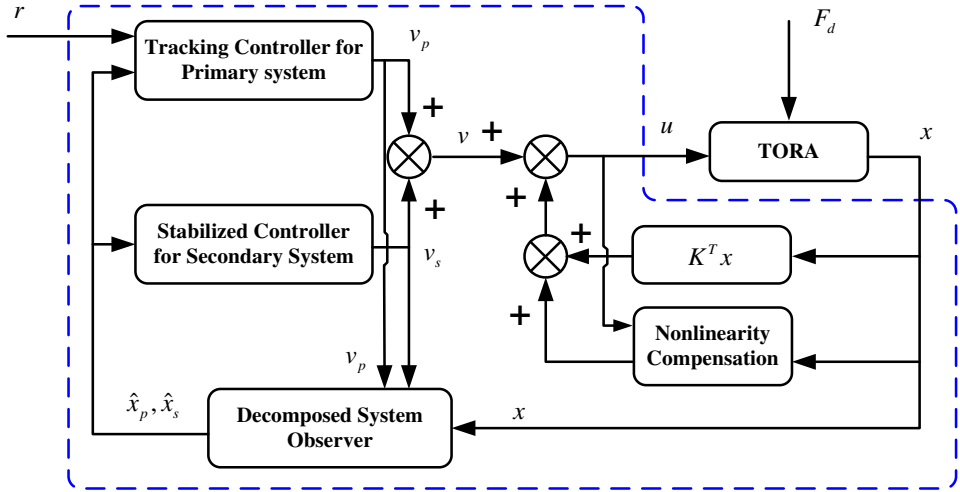}
\end{center}
\caption{Structure of the closed-loop system}%
\end{figure}

\subsection{Nonlinearity Compensation}

First, in order to estimate the term$\frac{\varepsilon\cos x_{3}%
}{1-\varepsilon^{2}\cos^{2}x_{3}}F_{d},$ an observer is designed, which is
stated in \textit{Theorem 1.}

\textit{Theorem 1}. Under \textit{Assumptions 1-2}, for system
(\ref{RTACmodel0}), let the observer be designed as follows%
\begin{align}
\dot{\hat{w}}  &  =S\hat{w}+l_{1}\frac{\varepsilon\cos x_{3}}{1-\varepsilon
^{2}\cos^{2}x_{3}}C_{d}\left(  \hat{x}_{4}-x_{4}\right) \nonumber\\
\dot{\hat{x}}_{4}  &  =-l_{2}\left(  \hat{x}_{4}-x_{4}\right)  -l_{1}%
\frac{\varepsilon\cos x_{3}}{1-\varepsilon^{2}\cos^{2}x_{3}}C_{d}^{T}\hat
{w}+u\label{observer}\\
\hat{F}_{d}  &  =l_{1}C_{d}^{T}\hat{w},\hat{w}\left(  0\right)  =0,\hat{x}%
_{4}\left(  0\right)  =0\nonumber
\end{align}
where $l_{1},l_{2}>0.$ Then $\underset{t\rightarrow\infty}{\lim}%
\frac{\varepsilon\cos x_{3}}{1-\varepsilon^{2}\cos^{2}x_{3}}\tilde{F}%
_{d}\left(  t\right)  =0,$ where $\tilde{F}_{d}\triangleq\hat{F}_{d}-F_{d}.$

\textit{Proof}. See \textit{Appendix A}. $\square$

By using the observer (\ref{observer}), the controller $u$ in
(\ref{RTACmodel0})\ is designed as follows%
\[
u=K^{T}x+v+\frac{\varepsilon\cos x_{3}}{1-\varepsilon^{2}\cos^{2}x_{3}}\hat
{F}_{d}%
\]
where $K\in%
\mathbb{R}
^{4}$ and $v\in%
\mathbb{R}
$ will be specified later. Then the system (\ref{RTACmodel0}) becomes
\begin{align}
\dot{x}_{1}  &  =x_{2}\nonumber\\
\dot{x}_{2}  &  =-x_{1}+\varepsilon\sin x_{3}+F_{d}\nonumber\\
\dot{x}_{3}  &  =x_{4}\nonumber\\
\dot{x}_{4}  &  =K^{T}x+v+\frac{\varepsilon\cos x_{3}}{1-\varepsilon^{2}%
\cos^{2}x_{3}}\tilde{F}_{d},x\left(  0\right)  =x_{0}. \label{RTACmodel_1}%
\end{align}

\subsection{Additive State Decomposition of Original System}

Introduce a zero term $\varepsilon D\left(  C+aB\right)  ^{T}x-\varepsilon
\left(  y+a\dot{y}\right)  =0\ $into the system (\ref{RTACmodel_1}), where
$a>0,$ $B=[%
\begin{array}
[c]{cccc}%
0 & 0 & 0 & 1
\end{array}
]^{T}$, $C=[%
\begin{array}
[c]{cccc}%
0 & 0 & 1 & 0
\end{array}
]^{T}\ $and $\left(  C+aB\right)  ^{T}x=y+a\dot{y}.$ Then the system
(\ref{RTACmodel_1}) becomes%
\begin{align}
\dot{x}  &  =Ax+Bv+\phi\left(  y,\dot{y}\right)  +DF_{d}+\varphi\nonumber\\
y  &  =C^{T}x,x\left(  0\right)  =x_{0} \label{RTACmodel_m}%
\end{align}
where%
\begin{align}
A_{0}  &  =\left[
\begin{array}
[c]{cccc}%
0 & 1 & 0 & 0\\
-1 & 0 & 0 & 0\\
0 & 0 & 0 & 1\\
0 & 0 & 0 & 0
\end{array}
\right]  ,A=A_{0}+BK^{T}+\varepsilon D\left(  C+aB\right)  ^{T},\nonumber\\
D  &  =\left[
\begin{array}
[c]{c}%
0\\
1\\
0\\
0
\end{array}
\right]  ,\phi\left(  y,\dot{y}\right)  =\left[
\begin{array}
[c]{c}%
0\\
\varepsilon\sin y-\varepsilon\left(  y+a\dot{y}\right) \\
0\\
0
\end{array}
\right]  ,\varphi=\left[
\begin{array}
[c]{c}%
0\\
0\\
0\\
\frac{\varepsilon\cos x_{3}}{1-\varepsilon^{2}\cos^{2}x_{3}}\tilde{F}_{d}%
\end{array}
\right]  . \label{RTACmodel_m_par}%
\end{align}
The additive state decomposition is ready to apply to the system
(\ref{RTACmodel_m}), for which the primary system is chosen to be an LTI
system including all external signals as follows%
\begin{align}
\dot{x}_{p}  &  =Ax_{p}+Bv_{p}+d+\varphi\nonumber\\
y_{p}  &  =C^{T}x_{p},x_{p}\left(  0\right)  =x_{0} \label{RTACprimary1}%
\end{align}
where $d=\phi\left(  r,0\right)  +DF_{d}.$ Then, according to the rule
(\ref{Dif_Sec_Sys}), the secondary system is derived from the original system
(\ref{RTACmodel_m}) and the primary system (\ref{RTACprimary1}) as follows%
\begin{align}
\dot{x}_{s}  &  =Ax_{s}+Bv_{s}+\phi\left(  y_{p}+y_{s},\dot{y}_{p}+\dot{y}%
_{s}\right)  -\phi\left(  r,0\right) \nonumber\\
y_{s}  &  =C^{T}x_{s},x_{s}\left(  0\right)  =0 \label{RTACsecondary1}%
\end{align}
where $v_{s}=v-v_{p}.$ According to (\ref{Gen_RelationDif}), we have%
\begin{equation}
x=x_{p}+x_{s}\ \text{and }y=y_{p}+y_{s}. \label{additiverelation}%
\end{equation}

\textit{Remark 3}. The pair $\left(  A_{0},B\right)  $ is uncontrollable,
while the pair $\left(  A_{0}+\varepsilon D\left(  C+aB\right)  ^{T},B\right)
$ is controllable. Therefore, there always exists a vector$\ K$ such that
$A=A_{0}+BK^{T}+\varepsilon D\left(  C+aB\right)  ^{T}$ is a stable matrix.

\textit{Remark 4}. If $y_{p}\equiv r$ and $\dot{y}_{p}\equiv0,$ then $\left(
x_{s},v_{s}\right)  =0$ is a zero equilibrium point of the secondary system
(\ref{RTACsecondary1}).

So far, the nonlinear nonminimum phase tracking system (\ref{RTACmodel_m})\ is
decomposed into two systems by the additive state decomposition, where the
external signal $d+\varphi$ is shifted to (\ref{RTACprimary1}) and the
nonlinear term $\phi\left(  \cdot\right)  $ is shifted to
(\ref{RTACsecondary1}). The strategy here is to assign the tracking
(rejection) task to the primary system (\ref{RTACprimary1}) and stabilization
task to the secondary system (\ref{RTACsecondary1}). More concretely, in
(\ref{RTACprimary1}) design $v_{p}$ to track $r$, and design $v_{s}$ to
stabilize (\ref{RTACsecondary1}). If so, by the relationship
(\ref{additiverelation}), $y$ can track $r.$ In the following, controllers
$v_{p}$ and $v_{s}$ are designed separately.

\subsection{Tracking Controller Design for Primary System}

Before proceeding further, we have the following preliminary result.

Consider the following linear system
\begin{align}
\dot{z}_{1}  &  =S_{z}z_{1}+A_{12}e_{z}\nonumber\\
\dot{z}_{2}  &  =A_{21}z_{1}+A_{22}z_{2}+d_{1}+\varphi_{1}\nonumber\\
e_{z}  &  =C_{e}^{T}z_{2}+d_{2}+\varphi_{2},z\left(  0\right)  =z_{0}
\label{Tempsys1}%
\end{align}
where $S_{z}\in%
\mathbb{R}
^{m_{1}\times m_{1}}$ is a marginally stable matrix, $A_{12}\in%
\mathbb{R}
^{m_{1}},$ $C_{e}\in%
\mathbb{R}
^{m_{2}},$ $A_{21}\in%
\mathbb{R}
^{m_{2}\times m_{2}},$ $A_{22}\in%
\mathbb{R}
^{m_{2}\times m_{1}},$ $z_{1}\in%
\mathbb{R}
^{m_{1}},$ $z_{2},d_{1},\varphi_{1}\in%
\mathbb{R}
^{m_{2}},$ $z=[%
\begin{array}
[c]{cc}%
z_{1}^{T} & z_{2}^{T}%
\end{array}
]^{T}\in%
\mathbb{R}
^{m_{1}+m_{2}}$ and $e_{z},d_{2},\varphi_{2}\in%
\mathbb{R}
.$

\textit{Lemma 1}. Suppose i) $\varphi_{i}\left(  t\right)  $ is bounded on
$\left[  0,\infty\right)  $ and$\ \underset{t\rightarrow\infty}{\lim
}\left\Vert \varphi_{i}\left(  t\right)  \right\Vert =0,i=1,2,$ ii) every
element of $d_{1}\left(  t\right)  ,d_{2}\left(  t\right)  $ are bounded on
$\left[  0,\infty\right)  $ and can be generated by $\dot{w}_{z}=S_{z}w_{z},$
$d_{z}=C_{z}^{T}w_{z}\ $with appropriate initial values, where $C_{z}\in%
\mathbb{R}
^{m_{1}},$ iii) the parameters in (\ref{Tempsys1}) satisfy%
\begin{equation}
\max\operatorname{Re}\lambda\left(  A_{z}\right)  <0,A_{z}=\left[
\begin{array}
[c]{cc}%
S_{z} & A_{12}C_{e}^{T}\\
A_{21} & A_{22}%
\end{array}
\right]  . \label{RTACcondition}%
\end{equation}
Then in (\ref{Tempsys1}) $\underset{t\rightarrow\infty}{\lim}e_{z}\left(
t\right)  =0,$ meanwhile keeping $z_{1}\left(  t\right)  $ and $z_{2}\left(
t\right)  $ bounded.

\textit{Proof.} See\textit{ Appendix B}. $\square$

Define a filtered tracking error to be
\begin{equation}
e_{p}=\tilde{y}_{p}+a\dot{\tilde{y}}_{p}=\left(  C+aB\right)  ^{T}x_{p}-r
\label{definition}%
\end{equation}
where $\tilde{y}_{p}=y_{p}-r,$ $\dot{r}=0$ and $a>0$. Let us consider the
tracking problem for the primary system (\ref{RTACprimary1}). With
\textit{Lemma 1} in hand, the design of $v_{p}$ is stated in \textit{Theorem
2}.

\textit{Theorem 2}. For the primary system (\ref{RTACprimary1}), let the
controller $v_{p}$ be designed as follows%
\begin{align}
\dot{\xi}  &  =S_{a}\xi+L_{1}e_{p}\nonumber\\
v_{p}\left(  \xi,x_{p},r\right)   &  =L_{2}^{T}x_{p}+L_{3}^{T}\xi
\label{controllerforpri}%
\end{align}
where $S_{a}=$diag$\left(  0,S\right)  ,$ $L_{1}\in%
\mathbb{R}
^{m+1},L_{2}\in%
\mathbb{R}
^{4}$ and $L_{3}\in%
\mathbb{R}
^{m+1}$ satisfy%
\begin{equation}
\max\operatorname{Re}\lambda\left(  A_{a}\right)  <0,A_{a}=\left[
\begin{array}
[c]{cc}%
S_{a} & L_{1}\left(  C+aB\right)  ^{T}\\
BL_{3}^{T} & A+BL_{2}^{T}%
\end{array}
\right]  . \label{conditioncontrollerforpri}%
\end{equation}
Then $\underset{t\rightarrow\infty}{\lim}y_{p}\left(  t\right)  =r\ $%
and$\ \underset{t\rightarrow\infty}{\lim}\dot{y}_{p}\left(  t\right)  =0$
meanwhile keeping $x_{p}\left(  t\right)  $ and$\ \xi\left(  t\right)  $ bounded.

\textit{Proof}. Incorporating the controller (\ref{controllerforpri}) into the
primary system (\ref{RTACprimary1}) results in%
\begin{align*}
\dot{\xi}  &  =S_{a}\xi+L_{1}e_{p}\\
\dot{x}_{p}  &  =\left(  A+BL_{2}^{T}\right)  x_{p}+BL_{3}^{T}\xi+d+\varphi\\
e_{p}  &  =\left(  C+aB\right)  ^{T}x_{p}-r
\end{align*}
where the definition (\ref{definition}) is utilized. Moreover, every element
of $d\ $and$\ r\ $can be generated by an autonomous system in the form
$\dot{w}_{a}=S_{a}w_{a},d_{a}=C_{a}^{T}w_{a}\ $with appropriate initial
values, where $C_{a}=[%
\begin{array}
[c]{cc}%
1 & C_{d}^{T}%
\end{array}
]^{T}.$ By \textit{Lemma 1},\textit{ }if (\ref{conditioncontrollerforpri})
holds, then $\underset{t\rightarrow\infty}{\lim}e_{p}\left(  t\right)  =0$
meanwhile keeping $x_{p}\left(  t\right)  $ and$\ \xi\left(  t\right)  $
bounded. It is easy to see from (\ref{definition}) that both $\tilde{y}_{p}$
and $\dot{\tilde{y}}_{p}$ can be viewed as outputs of a stable system with
$e_{p}$ as input. This means that $\tilde{y}_{p}$ and $\dot{\tilde{y}}_{p}$
are bounded if $e_{p}$ is bounded. In addition, $\underset{t\rightarrow\infty
}{\lim}\tilde{y}_{p}\left(  t\right)  =0\ $and$\ \underset{t\rightarrow\infty
}{\lim}\dot{\tilde{y}}_{p}\left(  t\right)  =0.$ $\square$

In most of cases, the controller parameters $L_{1},L_{2}$ and $L_{3}$ in
(\ref{controllerforpri}) can be always found. This is shown in the following proposition.

\textit{Proposition 1.} For any $S=-S^{T}$ without eigenvalues $\pm j,$ the
parameters
\begin{equation}
L_{1}=[%
\begin{array}
[c]{cc}%
1 & C_{d}^{T}%
\end{array}
]^{T},L_{2}=-\frac{1}{a}C-B-\frac{1}{a}H-\frac{1}{a}K,L_{3}=-\frac{1}{a}L_{1}
\label{parameters}%
\end{equation}
can always make $\max\operatorname{Re}\lambda\left(  A_{a}\right)  <0,$ where
$H=[%
\begin{array}
[c]{cccc}%
0 & \varepsilon & 0 & 1
\end{array}
]^{T}.$

\textit{Proof}. See\textit{ Appendix C}. $\square$

\textit{Remark 5}. \textit{Proposition 1 }in fact implies that, in the
presence of any external signal (except for the frequencies at $\pm1$), the
controller (\ref{controllerforpri}) with parameters (\ref{parameters}) can
always make $\underset{t\rightarrow\infty}{\lim}y_{p}\left(  t\right)
=r\ $and$\ \underset{t\rightarrow\infty}{\lim}\dot{y}_{p}\left(  t\right)  =0$
meanwhile keeping $x_{p}\left(  t\right)  $ and$\ \xi\left(  t\right)  $
bounded. In other words, the disturbance like $\sin t$ cannot be dealt with,
which is consistent with \cite{Lan(2006)}. If the external signal contains the
component with frequencies at $\pm1,$ then such a frequency component can be
chosen not to compensate for, i.e., $S_{a}$ in (\ref{controllerforpri}) will
not contain eigenvalues $\pm j$.

\subsection{Stabilized Controller Design for Secondary System}

So far, we have designed the tracking controller for the primary system
(\ref{RTACprimary1}). In this section, we are going to design the stabilized
controller for the secondary system (\ref{RTACsecondary1}). It can be
rewritten as
\begin{align}
\dot{x}_{1,s}  &  =x_{2,s}\nonumber\\
\dot{x}_{2,s}  &  =-x_{1,s}+\varepsilon\sin\left(  x_{3,s}+r\right)
-\varepsilon\sin r+g\nonumber\\
\dot{x}_{3,s}  &  =x_{4,s}\nonumber\\
\dot{x}_{4,s}  &  =K^{T}x_{s}+v_{s},x_{s}\left(  0\right)  =0
\label{RTACsecondary2}%
\end{align}
where $g=\varepsilon\sin\left(  y_{p}+x_{3,s}\right)  -\varepsilon\sin\left(
r+x_{3,s}\right)  -\varepsilon\left(  y_{p}+a\dot{y}_{p}-r\right)  .$ Our
constructive procedure has been inspired by the design in \cite{Jiang(2000)}.
We will start the controller design procedure from the marginally stable
$\left(  x_{1,s},x_{2,s}\right)  $-subsystem.

\textit{Step 1}. Consider the $\left(  x_{1,s},x_{2,s}\right)  $-subsystem of
(\ref{RTACsecondary2}) with $x_{3,s}$ viewed as the virtual control input.
Differentiating the quadratic function $V_{1}=\frac{1}{2}\left(  x_{1,s}%
^{2}+x_{2,s}^{2}\right)  $ results in%
\begin{equation}
\dot{V}_{1}=\varepsilon x_{2,s}\left[  \sin\left(  x_{3,s}+r\right)  -\sin
r\right]  +\varepsilon x_{2,s}g. \label{RTAC_dV1}%
\end{equation}
Guided by the state-feedback design \cite{Jiang(1998)}, we introduce the
following \textquotedblleft Certainty Equivalence\textquotedblright\ (CE)
based virtual controller%
\begin{equation}
x_{3,s}=-b\text{atan}x_{2,s}+x_{3,s}^{\prime}. \label{RTAC_Transfer}%
\end{equation}
Then%
\begin{align}
\dot{x}_{1,s}  &  =x_{2,s}\nonumber\\
\dot{x}_{2,s}  &  =-x_{1,s}+2\varepsilon\sin\left(  \frac{-b\text{atan}%
x_{2,s}}{2}\right)  \cos\left(  \frac{-b\text{atan}x_{2,s}+2r}{2}\right)
+g^{\prime} \label{Secondarysubsys}%
\end{align}
where%
\begin{equation}
g^{\prime}=\varepsilon\sin\left(  r-b\text{atan}x_{2,s}+x_{3,s}^{\prime
}\right)  -\varepsilon\sin\left(  r-b\text{atan}x_{2,s}\right)  +g.
\label{Secondarysubsysg}%
\end{equation}
In order to ensure $\cos\left(  \frac{-b\text{atan}x_{2,s}+2r}{2}\right)  >0,$
the parameter $b$ is chosen to satisfy $0<b<2\left(  1-2\left\vert
r\right\vert \left/  \pi\right.  \right)  .$ Since $r\in(-\pi\left/  2\right.
,\pi\left/  2\right.  )$ is a constant, $b$ always exists. The term CE is used
here because $x_{3,s}^{\prime}=0$ in (\ref{RTAC_Transfer})\ makes $\dot{V}%
_{1}$ in (\ref{RTAC_dV1}) negative semidefinite as $g\equiv0.$

\textit{Step 2}. We will apply backstepping to the $\left(  x_{3,s}^{\prime
},x_{4,s}\right)  $-subsystem and design a nonlinear controller $v_{s}$ to
drive $x_{3,s}^{\prime}$ to the origin. By the definition (\ref{RTAC_Transfer}%
), $x_{3,s}^{\prime}=x_{3,s}+b$atan$x_{2,s}.$ Then the time derivative of the
new variable $x_{3,s}^{\prime}$ is%
\begin{equation}
\dot{x}_{3,s}^{\prime}=x_{4,s}+\psi+b\frac{1}{1+x_{2,s}^{2}}g
\label{Secondarysubsys2}%
\end{equation}
where $\psi=b\frac{1}{1+x_{2,s}^{2}}\left[  -x_{1,s}+\varepsilon\sin\left(
x_{3,s}+r\right)  -\varepsilon\sin r\right]  .$ Define a new variable
$x_{4,s}^{\prime}$ as follows%
\begin{equation}
x_{4,s}^{\prime}=x_{3,s}^{\prime}+x_{4,s}+\psi. \label{Secondarysubsys2def1}%
\end{equation}
Then (\ref{Secondarysubsys2}) becomes%
\[
\dot{x}_{3,s}^{\prime}=-x_{3,s}^{\prime}+x_{4,s}^{\prime}+b\frac{1}%
{1+x_{2,s}^{2}}g.
\]
By the definition (\ref{Secondarysubsys2def1}), the time derivative of the new
variable $x_{4,s}^{\prime}$ is%
\begin{align*}
\dot{x}_{4,s}^{\prime}  &  =\dot{x}_{3,s}^{\prime}+\dot{x}_{4,s}+\dot{\psi}\\
&  =-x_{3,s}^{\prime}+x_{4,s}^{\prime}+b\frac{1}{1+x_{2,s}^{2}}g+K^{T}%
x_{s}+v_{s}+\dot{\psi},
\end{align*}
where%
\begin{align*}
\dot{\psi}  &  =-2bx_{2,s}\frac{1}{\left(  1+x_{2,s}^{2}\right)  ^{2}}\left[
-x_{1,s}+\varepsilon\sin\left(  x_{3,s}+r\right)  -\varepsilon\sin r\right] \\
&  \text{ \ \ }+b\frac{1}{1+x_{2,s}^{2}}\left[  -x_{2,s}+\varepsilon
\cos\left(  x_{3,s}+r\right)  x_{4,s}\right]  .
\end{align*}
Design $v_{s}$ for the secondary system (\ref{RTACsecondary2}) as follows%
\begin{equation}
v_{s}\left(  x_{p},x_{s},r\right)  =x_{3,s}^{\prime}-2x_{4,s}^{\prime}%
-K^{T}x_{s}-\dot{\psi}. \label{RTAC_controlforsecon}%
\end{equation}
Then the $\left(  x_{3,s}^{\prime},x_{4,s}^{\prime}\right)  $-subsystem
becomes%
\begin{align}
\dot{x}_{3,s}^{\prime}  &  =-x_{3,s}^{\prime}+x_{4,s}^{\prime}+b\frac
{1}{1+x_{2,s}^{2}}g\nonumber\\
\dot{x}_{4,s}^{\prime}  &  =-x_{4,s}^{\prime}+b\frac{1}{1+x_{2,s}^{2}}g.
\label{Secondarysubsys34}%
\end{align}
It is easy to see that $\underset{t\rightarrow\infty}{\lim}x_{3,s}^{\prime
}\left(  t\right)  =0$ and $\underset{t\rightarrow\infty}{\lim}x_{4,s}%
^{\prime}\left(  t\right)  =0\ $as $\underset{t\rightarrow\infty}{\lim
}g\left(  t\right)  =0.$

We are now ready to state the theorem for the secondary system.

\textit{Theorem 3}. Suppose $\underset{t\rightarrow\infty}{\lim}y_{p}\left(
t\right)  =r$ and $\underset{t\rightarrow\infty}{\lim}\dot{y}_{p}\left(
t\right)  =0.$ Let the controller $v_{s}$ for the secondary system
(\ref{RTACsecondary2}) be designed as (\ref{RTAC_controlforsecon}), where
$0<b<2\left(  1-2\left\vert r\right\vert \left/  \pi\right.  \right)  $. Then
$\underset{t\rightarrow\infty}{\lim}\left\Vert x_{s}\left(  t\right)
\right\Vert =0$ meanwhile keeping $x_{s}\left(  t\right)  $ bounded.

\textit{Proof. }See\textit{ Appendix D}. $\square$

\subsection{Controller Synthesis for Original System}

It should be noticed that the controller design above is based on the
condition that $x_{p}$ and $x_{s}$ are known as priori. A problem arises that
the states $x_{p}$ and $x_{s}$ cannot be measured directly except for
$x=x_{p}+x_{s}$. By taking this into account, the following observer is
proposed to estimate the states $x_{p}$ and $x_{s}$, which is stated in
\textit{Theorem 4.}

\textit{Theorem 4. }Let\textit{ }the observer be designed as follows%
\begin{align}
\dot{\hat{x}}_{s}  &  =A\hat{x}_{s}+Bv_{s}+\phi\left(  y,\dot{y}\right)
-\phi\left(  r,0\right) \nonumber\\
\hat{x}_{p}  &  =x-\hat{x}_{s},\hat{x}_{s}\left(  0\right)  =0
\label{systemobserver}%
\end{align}
where $A$ is stable. Then $\hat{x}_{p}\equiv x_{p}\ $and $\hat{x}_{s}\equiv
x_{s}.$

\textit{Proof. }Since $x=x_{p}+x_{s},$ we have $y=y_{p}+y_{s}.$ Consequently,
(\ref{systemobserver}) can be rewritten as%
\begin{align}
\dot{\hat{x}}_{s}  &  =A\hat{x}_{s}+Bv_{s}+\phi\left(  y_{p}+y_{s},\dot{y}%
_{p}+\dot{y}_{s}\right)  -\phi\left(  r,0\right) \nonumber\\
\hat{x}_{p}  &  =x-\hat{x}_{s},\hat{x}_{s}\left(  0\right)  =0.
\label{systemobserver1}%
\end{align}
Subtracting (\ref{RTACsecondary1})\ from (\ref{systemobserver1}) results in%
\begin{equation}
\dot{\tilde{x}}_{s}=A\tilde{x}_{s},\tilde{x}_{s}\left(  0\right)  =0
\label{error dynamics}%
\end{equation}
where $\tilde{x}_{s}=\hat{x}_{s}-x_{s}.$ Then $\hat{x}_{s}\equiv x_{s}.$
Furthermore, with the aid of\textit{ }the relationship $x_{p}=x-x_{s},$ we
have $\hat{x}_{p}\equiv x_{p}.\ \square$

\textit{Remark 6. }Unlike traditional observers, the proposed observer can
estimate the states of the primary system and the secondary system directly
rather than asymptotically or exponentially. This can be explained that,
although the initial value $x_{0}$ is unknown, the initial value of either the
primary system or the secondary system can be specified exactly, leaving an
unknown initial value for the other system. The measurement $x$ and parameters
may be inaccurate. In this case, it is expected that small uncertainties lead
to $\hat{x}_{p}\ $close to $x_{p}$ (or $\hat{x}_{s}\ $close to $x_{s}$). From
(\ref{error dynamics}), a\textit{ }stable matrix\textit{ }$A\ $can ensure a
small $\tilde{x}_{s}$ in the presence of small uncertainties.

\textit{Theorem 5}. Suppose that the conditions of \textit{Theorems 1-4 }hold.
Let the controller $u$ in the system (\ref{RTACmodel0}) be designed as follows%
\begin{align}
\dot{\xi}  &  =S_{a}\xi+L_{1}[\left(  C+aB\right)  ^{T}\hat{x}_{p}%
-r]\nonumber\\
u  &  =K^{T}x+v_{p}\left(  \xi,\hat{x}_{p},r\right)  +v_{s}\left(  \hat{x}%
_{p},\hat{x}_{s},r\right)  +\frac{\varepsilon\cos x_{3}}{1-\varepsilon^{2}%
\cos^{2}x_{3}}\hat{F}_{d} \label{RTACmodelcontroller}%
\end{align}
where $\hat{F}_{d}$ is given by (\ref{observer}), $\hat{x}_{p}$ and $\hat
{x}_{s}$ are given by (\ref{systemobserver}), $v_{p}\left(  \cdot\right)  $ is
defined in (\ref{controllerforpri}), and $v_{s}\left(  \cdot\right)  $ is
defined in (\ref{RTAC_controlforsecon}). Then $\underset{t\rightarrow\infty
}{\lim}y\left(  t\right)  =r$ meanwhile keeping $x\ $and $\xi$ bounded.

\textit{Proof.} Note that the original system (\ref{RTACmodel0}), the primary
system (\ref{RTACprimary1}) and the secondary system (\ref{RTACsecondary1})
have the relationship: $x=x_{p}+x_{s}$ and $y=y_{p}+y_{s}.$ With the
controller (\ref{RTACmodelcontroller}), for the primary system
(\ref{RTACprimary1}), $\underset{t\rightarrow\infty}{\lim}y_{p}\left(
t\right)  =r$ meanwhile keeping $x_{p}\ $and$\ \xi$ bounded$\ $by
\textit{Theorem 2}. On the other hand, for the secondary system
(\ref{RTACsecondary1}), we have $\underset{t\rightarrow\infty}{\lim}%
x_{s}\left(  t\right)  =0\ $meanwhile keeping $x_{s}$ bounded on $\left[
0,\infty\right)  $ by \textit{Theorem 3. }In addition,\textit{ Theorem 4
}ensures that $\hat{x}_{p}\equiv x_{p}\ $and $\hat{x}_{s}\equiv x_{s}.$
Therefore, $\underset{t\rightarrow\infty}{\lim}y\left(  t\right)  =r$
meanwhile keeping $x\ $and $\xi$ bounded. $\square$

\section{Numerical Simulation}

In the simulation, set $\varepsilon=0.2\ $and the initial value $x_{0}=[%
\begin{array}
[c]{cccc}%
0 & 0 & 0 & 0
\end{array}
]^{T}\ $in (\ref{RTACmodel0}). The unknown dimensionless disturbance $F_{d}%
$\ is generated by an autonomous LTI system (\ref{RTACdisturbance}) with the
parameters as follows%
\[
S=\left[
\begin{array}
[c]{cc}%
0 & 2\\
-2 & 0
\end{array}
\right]  ,C_{d}=[%
\begin{array}
[c]{cc}%
1 & 0
\end{array}
]^{T},w\left(  0\right)  =[%
\begin{array}
[c]{cc}%
0 & 0.02
\end{array}
]^{T}.
\]
The objective here is to design a controller $u$ such that the output
$y\left(  t\right)  =x_{3}\left(  t\right)  \rightarrow r=0.5$ as
$t\rightarrow\infty$ meanwhile keeping the other states bounded.

The parameters of the observer (\ref{observer}) are chosen as $l_{1}=l_{2}%
=10$. In (\ref{RTACmodel_m_par}), the parameters of $A$ are chosen as
$a=1\ $and$\ K=[%
\begin{array}
[c]{cccc}%
0 & -\varepsilon & -1 & -2
\end{array}
]^{T}.$ Then $\max$Re$\left(  \lambda\left(  A\right)  \right)  =-0.01<0.$
Since matrix $S$ does not possess the eigenvalues $\pm j,$ the parameters of
the tracking controller (\ref{controllerforpri}) of the primary system can be
chosen according to \textit{Proposition 1 }that $L_{1}=[%
\begin{array}
[c]{cc}%
1 & C_{d}^{T}%
\end{array}
]^{T},L_{2}=0_{4\times1}$ and $L_{3}=-L_{1}.$ These make $A_{a}$ in
(\ref{conditioncontrollerforpri}) satisfies $\max\operatorname{Re}%
\lambda\left(  A_{a}\right)  =-0.0084<0$. The parameter $b$ of the stabilized
controller (\ref{RTAC_controlforsecon}) is chosen as $b=1.5(1-1\left/
\pi\right.  )$ $<2\left(  1-2\left\vert r\right\vert \left/  \pi\right.
\right)  .$

The TORA system (\ref{RTACmodel0}) is driven by the controller
(\ref{RTACmodelcontroller}) with the parameters above. The evolutions of all
states of (\ref{RTACmodel0}) are shown in Fig.3. As shown, the proposed
controller $u$ drives the output $y\left(  t\right)  =x_{3}\left(  t\right)
\rightarrow0.5$ as $t\rightarrow\infty$, meanwhile keeping the other states bounded.

\begin{figure}[h]
\begin{center}
\includegraphics[
scale=0.65 ]{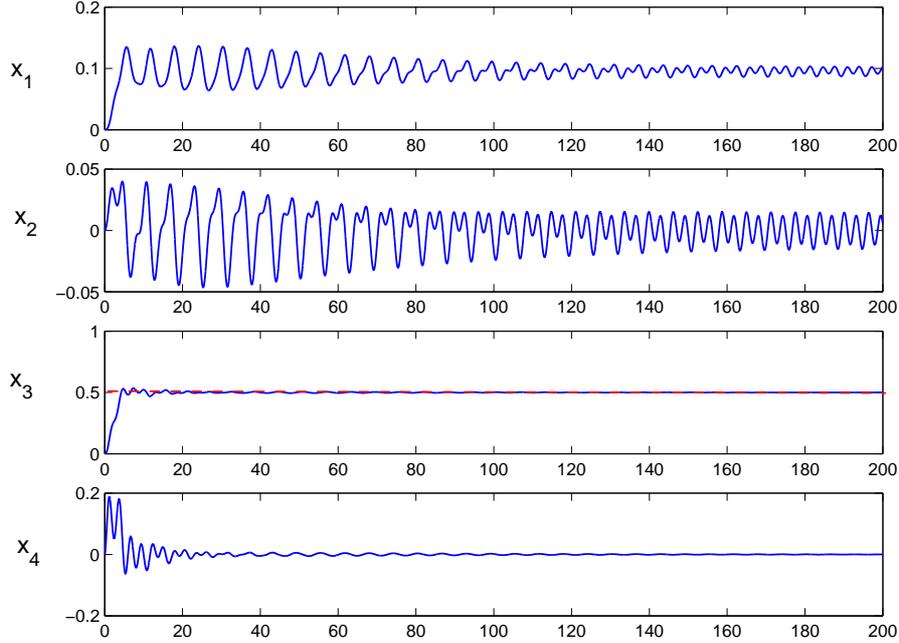}
\end{center}
\caption{Evolutions of all states}%
\end{figure}

Unlike the output regulation theory, the proposed method does not require the
regulator equations. If the disturbance $F_{d}$ consists of more frequency
components, i.e., $S$ is more complicated, the designed controller above does
not need to be changed except for the corresponding $S$ and $C_{d}$. This
demonstrates the effectiveness of the proposed control method. For example, we
consider that the unknown dimensionless disturbance $F_{d}$\ is generated by
an autonomous LTI system (\ref{RTACdisturbance}) with the parameters as
follows%
\begin{align*}
S  &  =\text{diag}\left(  S_{1},S_{2}\right)  ,S_{1}=\left[
\begin{array}
[c]{cc}%
0 & 2\\
-2 & 0
\end{array}
\right]  ,S_{2}=\left[
\begin{array}
[c]{cc}%
0 & 1.5\\
-1.5 & 0
\end{array}
\right] \\
C_{d}  &  =[%
\begin{array}
[c]{cccc}%
1 & 0 & 1 & 0
\end{array}
]^{T},w\left(  0\right)  =[%
\begin{array}
[c]{cccc}%
0 & 0.02 & 0 & 0.02
\end{array}
]^{T}.
\end{align*}
The controller in the first simulation is still applied to this case except
for replacing $S$ and $C_{d}$ (the dimension is changed correspondingly).
Driven by the new controller, the evolutions of all states of
(\ref{RTACmodel0}) are shown in Fig.4. As shown, the proposed controller $u$
drives the output $y\left(  t\right)  =x_{3}\left(  t\right)  \rightarrow0.5$
as $t\rightarrow\infty,$ meanwhile keeping the other states
bounded.\begin{figure}[h]
\begin{center}
\includegraphics[
scale=0.65 ]{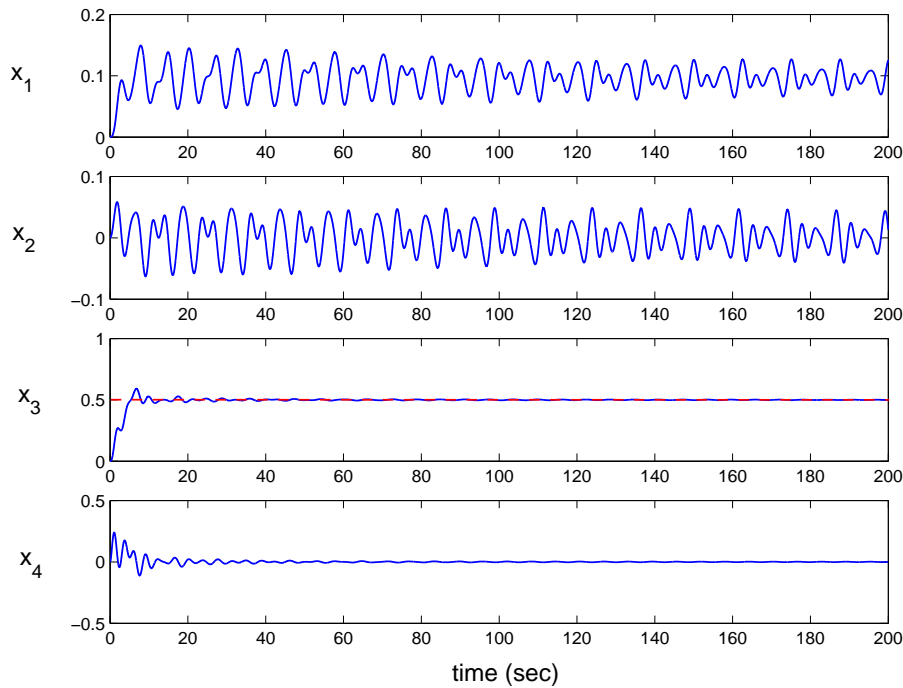}
\end{center}
\caption{Evolutions of all states when disturbance is complicated}%
\end{figure}

\section{Conclusions}

In this paper, the tracking (rejection) problem for rotational position of the
TORA was discussed. Our main contribution lies in the presentation of a new
decomposition scheme, named additive state decomposition, which not only
simplifies the controller design but also increases flexibility of the
controller design. By the additive state decomposition, the considered system
was decomposed into two subsystems in charge of two independent subtasks
respectively: an LTI system in charge of a tracking (rejection) subtask,
leaving a nonlinear system in charge of a stabilization subtask. Based on the
decomposition, the subcontrollers corresponding to two subsystems were
designed separately, which increased the flexibility of design. The tracking
(rejection) controller was designed by the transfer function method, while the
stabilized controller was designed by the backstepping method. This numerical
simulation has shown that the designed controller can achieve the objective,
moreover, can be changed flexibly according to the model of external signals.

\section{Appendix}

\subsection{Proof of Theorem 1}

The disturbance $F_{d}\in%
\mathbb{R}
$ is generated by an autonomous LTI system (\ref{RTACdisturbance}) with an
initial value $w\left(  0\right)  .$ It can also be generated by the following
system%
\begin{equation}
\dot{w}=Sw,F_{d}=l_{1}C_{d}^{T}w \label{RTACdisturbance1}%
\end{equation}
with the initial value $\frac{1}{l_{1}}w\left(  0\right)  .$ Subtracting
(\ref{RTACmodel4}) and (\ref{RTACdisturbance1})\ from (\ref{observer}) results
in%
\begin{align}
\dot{\tilde{w}}  &  =S\tilde{w}+l_{1}\frac{\varepsilon\cos x_{3}%
}{1-\varepsilon^{2}\cos^{2}x_{3}}C_{d}\tilde{x}_{4}\nonumber\\
\dot{\tilde{x}}_{4}  &  =-l_{2}\tilde{x}_{4}-l_{1}\frac{\varepsilon\cos x_{3}%
}{1-\varepsilon^{2}\cos^{2}x_{3}}C_{d}^{T}\tilde{w} \label{error sys}%
\end{align}
where $l_{1},l_{2}>0,$ $\tilde{x}_{4}\triangleq\hat{x}_{4}-x_{4}$ and
$\tilde{w}\triangleq\hat{w}-w.$ Design a Lyapunov function as follows%
\[
V_{1}=\frac{1}{2}\tilde{w}^{T}\tilde{w}+\frac{1}{2}\tilde{x}_{4}^{2}.
\]
Taking the derivative of $V_{1}$ along (\ref{error sys}) results in%
\begin{align*}
\dot{V}_{1}  &  =\frac{1}{2}\tilde{w}^{T}\left(  S+S^{T}\right)  \tilde
{w}+l_{1}\tilde{w}^{T}\frac{\varepsilon\cos x_{3}}{1-\varepsilon^{2}\cos
^{2}x_{3}}C_{d}\tilde{x}_{4}\\
&  -l_{2}\tilde{x}_{4}^{2}-l_{1}\tilde{x}_{4}\frac{\varepsilon\cos x_{3}%
}{1-\varepsilon^{2}\cos^{2}x_{3}}C_{d}^{T}\tilde{w}.
\end{align*}
By \textit{Assumption 2},\textit{ }$S+S^{T}=0.$ Then the derivative of $V_{1}$
becomes%
\[
\dot{V}_{1}\leq-l_{2}\tilde{x}_{4}^{2}\leq0.
\]
Since $l_{2}>0,$ from the inequality above, it can be concluded by LaSalle's
invariance principle \cite{Khalil (2002)} that $\underset{t\rightarrow\infty
}{\lim}\tilde{x}_{4}\left(  t\right)  =0$ and $\underset{t\rightarrow\infty
}{\lim}\frac{\varepsilon\cos x_{3}}{1-\varepsilon^{2}\cos^{2}x_{3}}C_{d}%
^{T}\tilde{w}\left(  t\right)  =0.$ $\square$

\subsection{Proof of Lemma 1}

Before proving \textit{Lemma 1}, we need the following preliminary result.

\textit{Lemma 2}. If the pair $\left(  A_{z},B_{z}\right)  $ is controllable,
then there exists a $C_{0}\in%
\mathbb{R}
^{m}$ such that
\[
C_{0}^{T}(sI_{m}-A_{z})^{-1}B_{z}=\frac{1}{\det\left(  sI_{m}-A_{z}\right)  }%
\]
where $A_{z}\in%
\mathbb{R}
^{m\times m}\ $and$\ B_{z}\in%
\mathbb{R}
^{m}.$

\textit{Proof. }First, we have%
\[
(sI_{m}-A_{z})^{-1}B_{z}=N[%
\begin{array}
[c]{ccc}%
s^{n-1} & \cdots & 1
\end{array}
]^{T}\left/  \det\left(  sI_{m}-A_{z}\right)  \right.
\]
where $N\in%
\mathbb{R}
^{m\times m}.$ If the pair $\left(  A_{z},B_{z}\right)  $ is controllable, the
matrix $N$ is full rank \cite{Cao (2008)}. We can complete this proof by
choosing $C_{0}=\left(  N^{-1}\right)  ^{T}[%
\begin{array}
[c]{cccc}%
0 & \cdots & 0 & 1
\end{array}
]^{T}.$ $\square$

With \textit{Lemma 2 }in hand, we are ready to prove \textit{Lemma 1}.

i) For the system (\ref{Tempsys1}), we have%
\[
z\left(  t\right)  =e^{-A_{z}t}z_{0}+%
{\displaystyle\int\nolimits_{0}^{t}}
e^{-A_{z}\left(  t-\tau\right)  }\left(  d_{a}+\varphi_{a}\right)  \left(
\tau\right)  d\tau,t\geq0
\]
where$\ d_{a}=[%
\begin{array}
[c]{cc}%
d_{2}^{T}A_{12}^{T} & d_{1}^{T}%
\end{array}
]^{T}$ and $\varphi_{a}=[%
\begin{array}
[c]{cc}%
\varphi_{2}^{T}A_{12}^{T} & \varphi_{1}%
\end{array}
]^{T}.$ Based on the equation above, since $\lambda\left(  A_{z}\right)  <0$
and $\left\Vert d_{a}\left(  t\right)  \right\Vert $, $\left\Vert \varphi
_{a}\left(  t\right)  \right\Vert $ are bounded on $\left[  0,\infty\right)
$, it is easy to see that $\left\Vert z_{1}\left(  t\right)  \right\Vert $ and
$\left\Vert z_{2}\left(  t\right)  \right\Vert $ are bounded on $\left[
0,\infty\right)  $.

ii) For the system (\ref{Tempsys1}), the Laplace transformation of $z\left(
s\right)  \ $is%
\[
z\left(  s\right)  =\left(  sI-A_{z}\right)  ^{-1}\left[  d_{a}\left(
s\right)  +\varphi_{a}\left(  s\right)  +z_{0}\right]  .
\]
Then $z_{1}\left(  s\right)  =C_{a}\left(  sI-A_{z}\right)  ^{-1}\left[
d_{a}\left(  s\right)  +\varphi_{a}\left(  s\right)  +z_{0}\right]  ,$ where
$C_{a}=[%
\begin{array}
[c]{cc}%
I_{m_{1}} & 0_{m_{1}\times m_{2}}%
\end{array}
]^{T}.$ The condition $\lambda\left(  A_{z}\right)  <0$ implies that the pair
$\left(  S_{z},A_{12}\right)  \ $is controllable. Otherwise, for the
autonomous system $\dot{z}=A_{z}z,$ the variable $z_{1}$ cannot converge to
zero as $S_{z}$ is a marginally stable matrix. This contradicts with the
condition $\lambda\left(  A_{z}\right)  <0$. Then by \textit{Lemma 1}, there
exists a $C_{0}\in%
\mathbb{R}
^{m_{1}}$ such that%
\[
C_{0}^{T}z_{1}\left(  s\right)  =\frac{1}{\det\left(  sI-S_{z}\right)  }%
e_{z}\left(  s\right)  .
\]
Then $e_{z}\left(  s\right)  $ can be written as%
\begin{align*}
e_{z}\left(  s\right)   &  =\det\left(  sI-S_{z}\right)  C_{0}^{T}z_{1}\left(
s\right) \\
&  =Q\left(  s\right)  \det\left(  sI-S_{z}\right)  \left[  d_{a}\left(
s\right)  +\varphi_{a}\left(  s\right)  +z_{0}\right]  .
\end{align*}
where $Q\left(  s\right)  =C_{0}^{T}C_{a}\left(  sI-A_{z}\right)  ^{-1}D_{a}.$
Since every element of $d_{a}$ can be generated by $\dot{w}_{z}=S_{z}%
w_{z},d_{z}=C_{z}^{T}w_{z}$, we have$\ d_{a}\left(  s\right)  =[C_{z}%
^{T}\left(  sI-S_{z}\right)  ^{-1}w_{z,i}\left(  0\right)  ]_{\left(
m_{1}+m_{2}\right)  \times1},$ where $w_{z,i}\left(  0\right)  \in%
\mathbb{R}
.$ Since $\left(  sI-S_{z}\right)  ^{-1}=\frac{1}{\det\left(  sI-S_{z}\right)
}$adj$\left(  sI-S_{z}\right)  $, $e_{z}\left(  s\right)  $ is further
represented as%
\begin{align}
e_{z}\left(  s\right)   &  =Q\left(  s\right)  C_{d}^{T}w_{z}\left(  0\right)
\left[  C_{z}^{T}\text{adj}\left(  sI-S_{z}\right)  w_{z,i}\left(  0\right)
\right]  _{\left(  m_{1}+m_{2}\right)  \times1}\nonumber\\
&  \text{ \ \ }+\left(  sI-A_{z}\right)  ^{-1}\left[  \varphi_{a}\left(
s\right)  +z_{0}\right]  . \label{error}%
\end{align}
Since $\lambda\left(  A_{z}\right)  <0\ $and the order of $A_{z}$ is higher
than that of $S_{z},$ moreover $\left\Vert \varphi_{a}\left(  t\right)
\right\Vert $ is bounded on $\left[  0,\infty\right)  $ and$\ \underset
{t\rightarrow\infty}{\lim}\left\Vert \varphi_{a}\left(  t\right)  \right\Vert
=0,$ for any initial value $w_{z,i}\left(  0\right)  ,$ we have $\underset
{t\rightarrow\infty}{\lim}e_{z}\left(  t\right)  =0$ from (\ref{error}).
$\square$

\subsection{Proof of Proposition 1}

If we can prove that the following system%
\begin{align}
\dot{\xi}  &  =S_{a}\xi+L_{1}\left(  C+aB\right)  ^{T}x_{p}\nonumber\\
\dot{x}_{p}  &  =\left(  A+BL_{2}^{T}\right)  x_{p}+BL_{3}^{T}\xi
\label{Primarysystemnodist}%
\end{align}
is asymptotic stable, then $\operatorname{Re}\lambda\left(  A_{a}\right)  <0$
holds. Choose a Lyapunov function as follows%
\[
V=\frac{1}{2}\xi^{T}\xi+\frac{1}{2}x_{1,p}^{2}+\frac{1}{2}x_{2,p}^{2}+\frac
{1}{2}p^{2}%
\]
where $p=x_{3,p}+ax_{4,p}=\left(  C+aB\right)  ^{T}x_{p}.$ With the parameters
$L_{1}=[%
\begin{array}
[c]{cc}%
1 & C_{d}^{T}%
\end{array}
]^{T},L_{2}=-\frac{1}{a}C-B-\frac{1}{a}H-K$ and $L_{3}=-\frac{1}{a}L_{1},$ the
derivative of $V$ along (\ref{Primarysystemnodist}) is%
\[
\dot{V}=-p^{2}.
\]
Define $\mathcal{S}=\left\{  \left.  x\right\vert \dot{V}\left(  x\right)
=0\right\}  ,$ where $x=[%
\begin{array}
[c]{cc}%
\xi^{T} & x_{p}^{T}%
\end{array}
]^{T}.$ The remaindering work is to prove $\mathcal{S}=\left\{  \left.
x\right\vert x=0\right\}  .$ If so, by LaSalle's invariance principle
\cite{Khalil (2002)}, we have $\underset{t\rightarrow\infty}{\lim}\left\Vert
x\left(  t\right)  \right\Vert =0.$ Therefore, the
system\ (\ref{Primarysystemnodist}) with the parameters is globally
asymptotically stable. Then $\operatorname{Re}\lambda\left(  A_{a}\right)  <0$.

Since $\dot{V}=0\Rightarrow p=0\ $and $a>0,$ we have $\mathcal{S}=\left\{
\left.  x\right\vert x_{3,p}=x_{4,p}=0\right\}  .$ Let $x$ be a solution
belonging to $\mathcal{S}$ identically. Then, from (\ref{Primarysystemnodist}%
), we have%
\begin{align}
\dot{\xi}  &  =S_{a}\xi\label{RTACequ1}\\
\dot{x}_{1,p}  &  =x_{2,p},\dot{x}_{2,p}=-x_{1,p}\label{RTACequ2}\\
0  &  =-\varepsilon x_{2,p}+L_{3}^{T}\xi\label{RTACequ3}%
\end{align}
From (\ref{RTACequ2}), it holds that
\[
x_{2,p}\in\mathcal{S}_{1}=\left\{  \left.  [%
\begin{array}
[c]{cc}%
0 & 1
\end{array}
]\xi\right\vert \dot{\xi}=\left[
\begin{array}
[c]{cc}%
0 & 1\\
-1 & 0
\end{array}
\right]  \xi\right\}  .
\]
On the other hand, from (\ref{RTACequ1}) and (\ref{RTACequ3}), it holds that
\[
x_{2,p}\in\mathcal{S}_{2}=\left\{  \left.  \frac{1}{\varepsilon}L_{3}^{T}%
\xi\right\vert \dot{\xi}=S_{a}\xi\right\}
\]
where matrix $S_{a}$ does not possess eigenvalues $\pm j\ $as matrix $S$ does
not. Therefore $x_{2,p}\in\mathcal{S}_{1}\cap\mathcal{S}_{2}=\left\{
0\right\}  \ $and then $x_{1,p}=0$, namely $\mathcal{S}=\left\{  \left.
x\right\vert x_{p}=0\right\}  .$

Let $x$ be a solution that belongs identically to $\mathcal{S}.$ Then
$L_{3}^{T}\xi=0.$ Since the pair $\left(  C_{d}^{T},S\right)  $ is observable,
by the definition $L_{1}=-\frac{1}{a}[%
\begin{array}
[c]{cc}%
1 & C_{d}^{T}%
\end{array}
]^{T},$ the pair $\left(  L_{3}^{T},S_{a}\right)  $ is observable as well.
Consequently, we can conclude that $\xi=0,$ namely $\mathcal{S}=\left\{
\left.  x\right\vert x=[%
\begin{array}
[c]{cc}%
\xi^{T} & x_{p}^{T}%
\end{array}
]^{T}=0\right\}  .$ $\square$

\subsection{Proof of Theorem 3}

This proof is composed of three parts.

\textit{Part 1. }$\underset{t\rightarrow\infty}{\lim}x_{3,s}^{\prime}\left(
t\right)  =0,$ $\underset{t\rightarrow\infty}{\lim}x_{4,s}^{\prime}\left(
t\right)  =0$ and $\underset{t\rightarrow\infty}{\lim}g^{\prime}\left(
t\right)  =0$ as $\underset{t\rightarrow\infty}{\lim}g\left(  t\right)  =0.$
If\textit{ }$\underset{t\rightarrow\infty}{\lim}y_{p}\left(  t\right)  =r$ and
$\underset{t\rightarrow\infty}{\lim}\dot{y}_{p}\left(  t\right)  =0,$ then
from the definition of $\phi\left(  y,\dot{y}\right)  ,$ we have
$\underset{t\rightarrow\infty}{\lim}g\left(  t\right)  =0\ $no matter what
$y_{s}$ is. According to this, it is easy from (\ref{Secondarysubsys34}) to
see that $\underset{t\rightarrow\infty}{\lim}x_{3,s}^{\prime}\left(  t\right)
=0\ $and $\underset{t\rightarrow\infty}{\lim}x_{4,s}^{\prime}\left(  t\right)
=0\ $when the controller $v_{s}$ for the secondary system
(\ref{RTACsecondary2}) is designed as (\ref{RTAC_controlforsecon}). Then, in
(\ref{Secondarysubsys}), $\underset{t\rightarrow\infty}{\lim}g^{\prime}\left(
t\right)  =0.$

\textit{Part 2.} $\underset{t\rightarrow\infty}{\lim}x_{1,s}\left(  t\right)
=0$ and $\underset{t\rightarrow\infty}{\lim}x_{2,s}\left(  t\right)  =0.$
Since $0<b<2\left(  1-2\left\vert r\right\vert \left/  \pi\right.  \right)  ,$
the derivative $V_{1}$ in (\ref{RTAC_dV1}) negative semidefinite when
$g^{\prime}\left(  t\right)  \equiv0,$ namely,%
\[
\dot{V}_{1}=2\varepsilon x_{2,s}\sin\left(  \frac{-b\text{atan}x_{2,s}}%
{2}\right)  \cos\left(  \frac{-b\text{atan}x_{2,s}+2r}{2}\right)  \leq0,
\]
where the equality holds at some time instant $t\geq0$ if and only if
$x_{2,s}\left(  t\right)  =0.$ By LaSalle's invariance principle \cite{Khalil
(2002)} that $\underset{t\rightarrow\infty}{\lim}x_{1,s}\left(  t\right)  =0$
and $\underset{t\rightarrow\infty}{\lim}x_{2,s}\left(  t\right)  =0\ $when
$g^{\prime}\left(  t\right)  \equiv0.$ Because of the particular structure of
$\left(  x_{1,s},x_{2,s}\right)  $-subsystem (\ref{Secondarysubsys}), by using
\cite[Lemma 3.6]{Hkctor (1994)}, one can show that any globally asymptotically
stabilizing feedback for $\left(  x_{1,s},x_{2,s}\right)  $-subsystem
(\ref{Secondarysubsys}) when $g^{\prime}\left(  t\right)  \equiv0$ achieves
global asymptotic stability of $\left(  x_{1,s},x_{2,s}\right)  $-subsystem
(\ref{Secondarysubsys}) when $\underset{t\rightarrow\infty}{\lim}g^{\prime
}\left(  t\right)  =0$. Therefore, based on \textit{Part 1}, $\underset
{t\rightarrow\infty}{\lim}x_{1,s}\left(  t\right)  =0$ and $\underset
{t\rightarrow\infty}{\lim}x_{2,s}\left(  t\right)  =0.$

\textit{Part 3.}Combining the two parts above, we have $\underset
{t\rightarrow\infty}{\lim}\left\Vert x_{s}\left(  t\right)  \right\Vert =0.$
For the $\left(  x_{1,s},x_{2,s}\right)  $-subsystem and $\left(
x_{3,s}^{\prime},x_{4,s}^{\prime}\right)  $-subsystem, $\left\Vert
x_{s}\left(  t\right)  \right\Vert $ is bounded in any finite time. With the
obtained result $\underset{t\rightarrow\infty}{\lim}\left\Vert x_{s}\left(
t\right)  \right\Vert =0,$ we have $\left\Vert x_{s}\left(  t\right)
\right\Vert $ is bounded on $\left[  0,\infty\right)  $. $\square$

\end{document}